\documentclass[aps,twocolumn,prb,amssymb,floatfix]{revtex4}

\usepackage{graphicx,amsmath,epstopdf}

\begin{document}

\title{Doping front instabilities in organic semiconductors: a means for optimizing optoelectronic devices }

\author{V. Bychkov, O. Yukhimenko, M. Modestov, and M. Marklund}

\affiliation{Department of Physics, Ume{\aa} University,
SE-90187 Ume{\aa}, Sweden}


\begin{abstract}
Recently, it was demonstrated that electrochemical doping fronts in organic semiconductors exhibit a new fundamental instability growing from multidimensional perturbations [Phys. Rev. Lett. \textbf{107} 016103 (2011)]. In the instability development, linear growth of tiny perturbations goes over into a nonlinear stage of strongly distorted doping fronts.
Here we  develop the nonlinear theory of the doping front instability and predict the key parameters of a distorted doping front, such as its velocity, in close agreement with the experimental data. We show that the instability makes the electrochemical doping process considerably faster.
We  obtain the self-similar  properties of the front shape corresponding to the maximal propagation velocity, which allows for a wide range of controlling the doping process in the experiments.
The theory developed  provides the guide for optimizing the performance of organic optoelectronic devices.
\end{abstract}

\maketitle

\section{Introduction}

The discovery and development of $\pi$-conjugated polymers -- plastics with properties of semiconductors, often termed organic semiconductors (OSCs) -- has revolutionized the field of electronics, and was awarded with the Nobel Prize in Chemistry in 2000 \cite{Chiang,Malliaras,Heeger}. The OSCs, in contrast to crystalline inorganic semiconductors, are simple to process and relatively inexpensive in production, while  their soft and conformable character enables large number of novel technological applications ranging from optoelectronics and photovoltaics to sensors and artificial muscles \cite{Pei-et-al-1995,Sirringhaus,Forrest-2004,Morteani-et-al-2003,Leger-2008,Arkipov-et-al-2005}. The unique combination of electronic and ionic conductivity in OSCs introduces the exciting functionality that triggers experimental and technological activities in the nascent field of so-called organic (i.e. based on OSCs) electronics. At the same time, as formulated in Ref. \cite{Leger-2008}, "the understanding of the fundamental processes that take place [in OSCs] is still in its infancy", especially if  compared to the theoretical fundament of the field of crystalline  inorganic semiconductors \cite{Haas-et-al-2009,Morandi-et-al-2010}.  The inadequate understanding of fundamental OSC properties limits the technologies to  "cut-and-try" methods and demands development of the theory on the subject. In line with this demand, the goal of the present work is to attain better fundamental understanding of the doping process in OSCs, which may provide a guide to  optimizing performance of optoelectronic devices such as, for example, light-emitting electrochemical cells.
\begin{figure}
\centering
\includegraphics[width=3.2in]{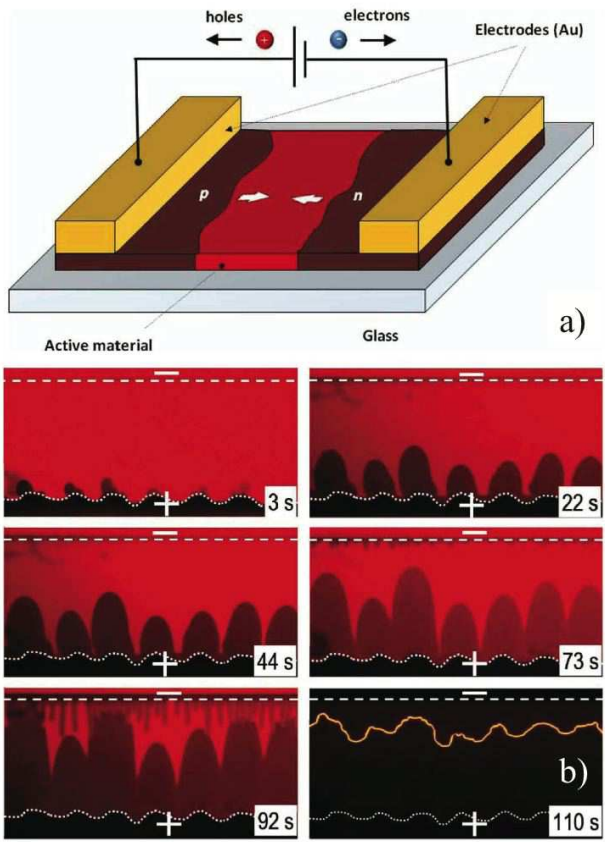}
\caption{Schematic of a light-emitting electrochemical cell (a). Experimental photographs of Ref. \cite{Bychkov-et-al-2011} showing the time evolution
of the doping fronts in a light-emitting electrochemical cell with the
corrugations imposed at the anode surface (b). The corresponding time instances are
indicated on the photographs.}
\end{figure}

In the process of electrochemical doping the OSCs can be transformed from an almost insulating state to a metallic-like state with the conductivity increased by several orders of magnitude
\cite{Pei-et-al-1995,Slinker-et-al-2007,Matyba-et-al-2009}. Apart from conductivity, this doping transformation also allows for tuning other key properties of OSCs, such as color, volume, surface energy and photoluminescence capability, in a controllable way. This possibility   can be profitably employed in numerous applications, e.g., in optoelectronics \cite{Heeger,Pei-et-al-1995,Sirringhaus,Forrest-2004,Morteani-et-al-2003,Leger-2008}.
Operationally, the electrochemical doping can be performed \textit{in-situ} by applying a potential difference between two metal electrodes in contact with an OSC, blended with a solid-state electrolyte, as illustrated in Fig. 1 (a). Because of the applied electric field, positive and negative ions redistribute in the active material and allow the injection of
 electrons (n-doping) and holes (p-doping) into the OSC from the cathode  and anode, respectively \cite{Pei-et-al-1996,Gao-Dane-2004,Matyba-et-al-2008,Lin-et-al-2010,Matyba-et-al-2010}.
As the electronic charges populate the OSC, its conductivity  increases by several orders of magnitude in comparison with the initial conductivity supported by ions. It has been demonstrated experimentally that the doping process develops in the form of two counter propagating fronts illustrated in Fig. 1 (a), which separate the regions of p- and n-doped material from the undoped one \cite{Matyba-et-al-2008,Matyba-et-al-2010,Bychkov-et-al-2011}. When the two fronts meet,  a p-n junction is formed in which the recombination of holes and electrons results in the visible light \cite{Pei-et-al-1995,Matyba-et-al-2008,Matyba-et-al-2010,Bychkov-et-al-2011,Hoven-et-al-2010,Smela-2008}.
Thus, the velocity of doping front propagation determines the turn-on time of the light-emitting electrochemical cells.

As pointed out above, the present understanding of the doping process in OSCs is far from being adequate, with only  few theoretical papers focusing on the subject
\cite{Bychkov-et-al-2011,Miomandre,Modestov-et-al-2010,vanReenen-11,Modestov-et-al-2011}. In particular, the analytical formula for the velocity of a planar doping front has been found only recently in Ref. \cite{Modestov-et-al-2011}, which demonstrated good agreement with the experimental observations. The analytical theory has shown that the doping rate is controlled by the mobility of heavy ions that results in a slow turn-on time of light-emitting electrochemical cells. At present, one of the most difficult problems to solve is to speed-up the doping process and to reduce the turn-on time. The speed-up could be achieved with the help of a new basic phenomenon -- the multidimensional doping front instability -- discovered recently in OSCs in Ref. \cite{Bychkov-et-al-2011}.
This new fundamental instability distorts the shape of doping fronts, increases locally the electric field ahead of the fronts and hence increases their propagation velocity. The recent paper \cite{Bychkov-et-al-2011} has developed a linear theory and provided the experimental evidence of the instability; it has also proposed a possible way to amplify the instability by means of minor humps introduced at the electrode surfaces.
Figure 1 (b) illustrates the strongly corrugated doping fronts obtained in the experiments of Ref. \cite{Bychkov-et-al-2011}. Still, the dynamics of the strong instability at the nonlinear stage remained unclear and demanded deep theoretical studies.

In the present paper we develop the nonlinear theory of the doping front instability, focusing on the velocity and shape of the corrugated fronts. For the 2D geometry of the OSC films, the theory predicts increase of the doping front velocity by a factor of two in comparison with the planar front velocity, which is   in a good agreement with the previous experimental data. We obtain self-similar properties of the front shape corresponding to the maximal velocity, which promises a wide range of controlling the doping process in the devices, e.g., through the design of the electrodes. We show that moderately corrugated doping fronts are preferential for achieving higher  velocity as compared with the strong corrugations obtained in Ref. \cite{Bychkov-et-al-2011} and illustrated in Fig. 1.
The developed theory provides a guide for optimization of organic optoelectronic devices utilizing electrochemical doping.

In addition, it is expected that doping fronts in OSCs will become a new popular object of nonlinear science alongside with other fronts and interfaces in fluids, gases and plasmas \cite{Cross-Hohenberg-review,Malomed-2000,Mendonca-2012}. In particular, in the present work we discuss similarity and difference between the doping front instability in OSCs and the hydrodynamic instability of a flame, i.e. a slow combustion front \cite{Zeldovich-et-al-1985,Law-book,Bychkov-Liberman-2000}. The electrochemical doping process provides a fascinating opportunity to observe directly and investigate a number of interesting nonlinear phenomena: front propagation, multi-dimensional instabilities and pattern formation, with surprisingly different outcome for p- and n-fronts. In addition, the present analysis suggests different regimes of the doping front instability for the 2D geometry of  OSC films and for the 3D geometry of the bulk optoelectronic devices \cite{Matyba-et-al-2010}: thus, the nonlinear dynamics of the instability depends also on the system dimension.

The paper is organized as follows: In Sec. II we remind  basic results on the linear stage of the doping front instability needed for the subsequent nonlinear analysis. In Sec. III we derive the nonlinear equation for the p-doping front bent by the instability. In Section IV we analyze stationary periodic solutions to the nonlinear equation. In Sec. V we compare the theoretical results to the experimental data and discuss technical implications of the theoretical predictions.  The paper is concluded by a brief summary in Sec. VI.

\section{Linear stage of the doping front instability}

In this section we remind basic results of Ref. \cite{Bychkov-et-al-2011} on the  doping front instability at the linear stage, when perturbation amplitude is infinitesimal. Details of the linear analysis are needed to understand the nonlinear theory, which we develop below in Sec. III and IV.
The instability arises because speed of a doping front depends on the electrical field ahead of the front as it was derived in \cite{Modestov-et-al-2011}.  We present the p/n-doping front velocity in the vector form as
\begin{equation}
\label{eq1-instab}
{\rm {\bf U}}_{p,n} = \pm {\frac{{n_{0}}} {{n_{h,e}}} }(\mu _{ +}  + \mu _{
-}  ){\rm {\bf E}} .
\end{equation}
Here ${\rm {\bf E}} = - \nabla \phi  $  is the electric field intensity just ahead of the front,  $\phi  $ the electric field potential, $n_{0} $  the initial ion concentration in the  active material, $n_{h,e} $  the final concentration of the holes/electrons behind the front determined by the thermodynamic properties of a particular OSC, $\mu _{ +}  $ and $\mu _{ -}  $  the mobilities of the positive and negative ions, respectively.
The "$\pm$" sign in Eq. (\ref{eq1-instab}) indicates that the p-type ($+$) and n-type ($-$) doping fronts propagate in opposite directions. We start our analysis with a single p-doping front propagating stationary with velocity $U_{p}$ in a constant uniform external electric field ${\bf E} = {\bf \hat{a}}_{x}E_{0}$ pointing along x-axis, where ${\bf \hat{a}}_{x}$ is the unit vector in x-direction. The analysis of the n-front instability may be performed in a similar way.
In the case of a planar stationary front, the general 2D equation for the p-front position $x = X_{p} (t,y)$ reduces  to $X_{p}(t) = U_{p}t$. Next, we consider small perturbations  of the stationary p-doping front
\begin{equation}
\label{perturbations}
X_{p}(t, y) = U_{p} t+\tilde {X}_{p}(t, y),
\end{equation}
where $\tilde {X}_{p}\propto\exp (\sigma t + iky)$ stands for one Fourier mode at the linear instability stage, $k= 2\pi/\lambda$ is the perturbation wave number characterizing infinitesimal bending of the doping front ($\lambda$ is the perturbation wavelength), $\sigma$ is the instability growth rate, see Fig. 2. The purpose of the linear stability analysis is to find the dispersion relation $\sigma(k)$. In the case of $\textrm{Re}(\sigma)>0$ valid for a certain domain of perturbation wave numbers we conclude that the planar front is unstable with respect to multidimensional perturbations.
\begin{figure}
\centering
\includegraphics[width=3.2in]{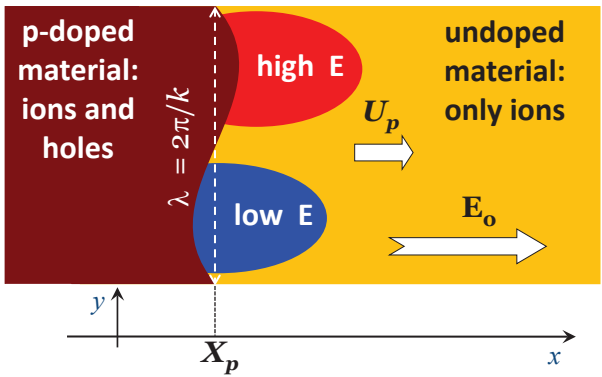}
\caption{Schematic of the doping front instability.}
\end{figure}
Small bending of the doping front induces also perturbations of the electric field in the undoped region $\tilde {\phi}(t,y,x)\propto \exp (\sigma t + iky)$.
The doped region may be treated as equipotential due to high conductivity.
Thickness of the doping front may be characterized by a length scale  $L_{p} $, of about $10^{ - 4}{\rm -}  10^{ - 3} {\rm m}{\rm m}$, determined by ionic diffusion \cite{Modestov-et-al-2010,Modestov-et-al-2011}. Since the characteristic size of the experimentally observed p-front perturbations (approximately $0.2 \textrm{mm}$) is much larger than  $L_{p} $, we may treat the front as infinitesimally thin.
Most of the time the doping fronts are sufficiently far away from each other,  $k(X_{n} - X_{p})\gg 1$, therefore, the instability of one front is not affected by the other.
The linearized form of Eq. (\ref{eq1-instab}) may be written in components as
\begin{equation}
\label{eq2a-instab}
\frac{\partial \tilde {X}_{p}}{\partial t} = - {\frac{{n_{0}}} {{n_{h}}} }(\mu _{ +}  + \mu
_{ -}  ) \frac{\partial \tilde {\phi}}{\partial x} ,
\end{equation}
\begin{equation}
\label{eq2b-instab}
ikU_{p} \tilde {X}_{p} = {\frac{{n_{0}}} {{n_{h}}} }(\mu _{ +}  + \mu _{ -}
)\frac{\partial \tilde {\phi}}{\partial y}  .
\end{equation}
Solving the Laplace equation $\nabla^{2}{\phi}=0$ for perturbations of the electric potential in the udoped region
\begin{equation}
\label{eq-Laplace-linear}
\frac{\partial^{2} \tilde {\phi}}{\partial x^{2}}-k^{2}\tilde {\phi}=0,
\end{equation}
we find the solution
\begin{equation}
\label{solution-Laplace-linear}
\tilde {\phi}  \propto \exp (\sigma t + iky - {\left|
{k} \right|}x).
\end{equation}
 Eqs. (\ref{eq2a-instab}), (\ref{eq2b-instab}), (\ref{solution-Laplace-linear}) may be reduced to a single linear equation describing the perturbation growth with time
\begin{equation}
\label{dispersion-relation}
\frac{\partial \tilde {X}_{p}}{\partial t} = U_{p} {\left| {k} \right|}\tilde {X}_{p}.
\end{equation}
 Thus, we obtain an instability for the p-type doping front, where the initially small perturbations grow exponentially as  $\tilde {X}_{p} \propto \exp(\sigma t)$, with the growth rate
 \begin{equation}
\label{growth rate}
\sigma = U_{p} {\left| {k} \right|}.
\end{equation}
 The same result holds for the n-type doping when $U_{p}$ is replaced with  $U_{n}$. The obtained dispersion relation  $\sigma \propto {\left| {k} \right|}$ is mathematically similar to the Darrieus-Landau (DL) instability encountered in combustion, astrophysics and laser fusion \cite{Zeldovich-et-al-1985,Law-book,Bychkov-Liberman-2000,Modestov-et-al-2009}. The role of the electric field in the new instability may be also compared to that of the gas velocity field in flames.

According to the above analysis, the doping front instability  develops because of a positive feedback of two effects: First, the local doping speed depends on the electric field, Eq. (\ref{eq1-instab}); and, second, the electric field increases close to the humps  at a curved  surface of the conducting material \cite{Landau-Lif-El}. The latter is reminiscent  of  the St. Elmo's fire phenomenon, which visualizes strong increase of electric field at sharp conducting surfaces by numerous discharges in the surrounding air. In the present configuration of a curved doping front, any small hump at the front produces local increase of the electric field in the undoped material close to the hump, which, in turn, facilitates further growth of the hump, thus leading to the instability.

\section{The nonlinear equation for the p-doping front bent by the instability}

A smooth shape of the p-doping front observed in the experiments \cite{Bychkov-et-al-2011} allows reducing the full diffusion-mobility model \cite{Modestov-et-al-2011} to a single nonlinear equation. Similar to the previous section, we consider a single p-doping front propagating in the initially uniform external electric field $\textbf{E}= {\rm {\bf \hat {a}}}_{x} E_{0}$.
In developing the nonlinear theory of the doping front instability for perturbations of considerable amplitudes,
we have to take into account interaction
of modes with different wave numbers. Then Eq. (\ref{dispersion-relation}) has to be presented in the form
\begin{equation}
\label{dispersion-relation-nonlin}
\frac{\partial \tilde {X}_{p}}{\partial t} = U_{p} \hat {J}\tilde {X}_{p},
\end{equation}
where the DL operator $\hat {J}$ implies product by ${\left| {k}
\right|}$ in Fourier space, i.e.
\begin{equation}
\label{DL-operator}
\hat {J}\tilde {X}_{p}(t,y) = {\int\limits_{ -
\infty} ^{\infty}  {{\left| {k} \right|}\tilde {X}_{pk}}}(t,k)  \exp (iky)dk
\end{equation}
in
the 2D geometry of an OSC film. Equation (\ref{dispersion-relation-nonlin}) will be obtained below as the  principal linear terms in the nonlinear equation for the curved p-doping front.

We study dynamics of  the p-doping front
 as a geometrical surface
 \begin{equation}
\label{G-surface}
G(\textbf{r},t)=x - U_{p} t-\tilde {X}_{p}(t, y)=0
\end{equation}
 using the identity
\begin{equation}
\label{G-equation}
\frac{dG}{dt}=\frac{d\textbf{r}}{dt}\cdot\nabla G + \frac{\partial G}{\partial t}=0,
\end{equation}
which leads to the normal  front velocity \cite{Law-book,Bychkov-Liberman-2000}
\begin{equation}
\label{u-normal}
{\rm {\bf \hat {a}}}_{n} \cdot \frac{d\textbf{r}}{dt} = -\frac{1}{\left|\nabla G \right|}\frac{\partial G}{\partial t}={\frac{{U_{p} + \partial _{t} \tilde {X}_{p}}}{{\sqrt
{1 + (\partial _{y} \tilde {X}_{p})^{2}}}} }.
\end{equation}
Here we have introduced the unit normal vector to the surface
\begin{equation}
\label{normal vector}
{\rm {\bf \hat {a}}}_{n} \equiv \frac{\nabla G}{\left|\nabla G \right|}=
\frac{{{\rm {\bf \hat {a}}}_{x} - {\rm {\bf \hat {a}}}_{y} \partial _{y} \tilde
{X}_{p}}}{{\sqrt {1 + (\partial _{y}
\tilde {X}_{p})^{2}}}}.
\end{equation}
On the other hand, the normal velocity of the curved p-doping front is determined by the electric field just ahead of the front, see Eq. (\ref{eq1-instab}), which leads to
\begin{equation}
\label{M1}
{\rm {\bf \hat {a}}}_{n} {\frac{{U_{p} + \partial _{t} \tilde {X}_{p}}}{{\sqrt
{1 + (\partial _{y} \tilde {X}_{p})^{2}}}} } = {\frac{{n_{0}}} {{n_{h}}} }(\mu _{ +}  + \mu _{ -}  ) {\rm ({\bf \hat {a}}}_{x} E_{0} -
\nabla \tilde {\phi
}).
\end{equation}
Taking into account the relation between the planar front velocity $U_{p}$ and the external electric field $E_{0}$, we write Eq. (\ref{M1}) in components as
\begin{equation}
\label{M2}
{\frac{{U_{p} + \partial _{t} \tilde {X}_{p}}}{{1 + (\partial _{y} \tilde
{X}_{p})^{2}}}} = U_{p} - {\frac{{U_{p}}} {{E_{0}}} }\;\partial _{x} \tilde {\phi} ,
\end{equation}
\begin{equation}
\label{M3}
- \partial _{y} \tilde {X}_{p}{\frac{{U_{p} + \partial _{t} \tilde {X}_{p}}}{{1 +
(\partial _{y} \tilde {X}_{p})^{2}}}} = - {\frac{{U_{p}}} {{E_{0}}} }\;\partial _{y} \tilde {\phi} .
\end{equation}
We also employ the solution to the Laplace equation for one Fourier mode of the perturbed electric field $\tilde {\phi}  \propto
\exp (iky - {\left| {k} \right|}x)$ and find the general identity
\begin{equation}
\label{operators}
\partial _{x} \tilde {\phi}  =
- \hat {J}\tilde {\phi}  = - \hat {J}(\partial _{y} \tilde {\phi}  / ik) = -
\hat {H}\partial _{y} \tilde {\phi} ,
\end{equation}
where the Hilbert operator $\hat {H}$
means multiplication by ${\left| {k} \right|} / ik$ in Fourier space. Then
Eqs. (\ref{M2}), (\ref{M3}) may be reduced to
\begin{equation}
\label{M4}
{\frac{{U_{p}^{ - 1} \partial _{t} \tilde {X}_{p} - (\partial _{y} \tilde
{X}_{p})^{2}}}{{1 + (\partial _{y} \tilde {X}_{p})^{2}}}} = \hat {H}{\left[
{\partial _{y} \tilde {X}_{p}{\frac{{1 + U_{p}^{ - 1} \partial _{t} \tilde
{X}_{p}}}{{1 + (\partial _{y} \tilde {X}_{p})^{2}}}}} \right]}.
\end{equation}
 Within the limit of weak second-order nonlinearity we simplify Eq. (\ref{M4}) as
\begin{equation}
\label{eq4-instab-preliminary}
U_{p}^{ - 1}\frac{\partial \tilde {X}_{p}}{\partial t} = \hat {J}\tilde {X}_{p} + \left(\frac{\partial \tilde {X}_{p}}{\partial y}\right)^{2} + U_{p}^{ - 1} \hat {H}{\left[ \frac{\partial \tilde {X}_{p}}{\partial t}\;\frac{\partial \tilde {X}_{p}}{\partial y} \right]}.
\end{equation}
The first two terms in Eq. (\ref{eq4-instab-preliminary}) describe the linear instability growth similar to Eq. (\ref{dispersion-relation-nonlin}). The third term corresponds to the nonlinear Huygens  stabilization of the instability due to the  front propagation, which is well-known, e.g., in combustion science \cite{Zeldovich-et-al-1985,Law-book,Bychkov-Liberman-2000}.
Because of the Huygens  stabilization convex parts of the front (humps) become smoother, while concave parts converge to sharp cusps, as illustrated in Fig. 3. Propagation speed of the cusps increases in comparison with the planar front velocity $U_{p}$, and the cusps tend to catch up with the humps. When linear growth of the humps because of the instability is balanced by the nonlinear increase of the cusp velocity, we should expect formation  of a curved stationary or quasi-stationary p-doping front propagating with an increased velocity in comparison with $U_{p}$. Still, one should employ the similarity between  combustion fronts and the doping fronts in OSCs with caution, since these two phenomena demonstrate also a number of important different features apart from  similar ones.
\begin{figure}
\centering
\includegraphics[width=2.5in]{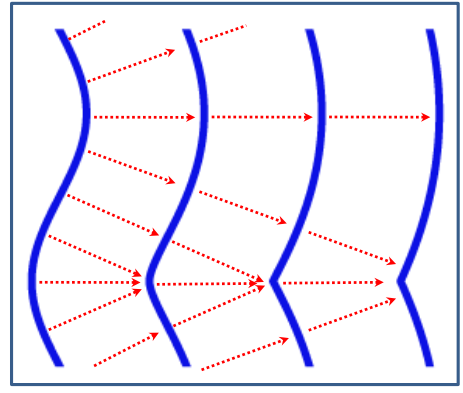}
\caption{Schematic of the Huygens nonlinear stabilization: solid lines represent successive front positions; dotted arrows indicate local direction of front propagation.}
\end{figure}
In particular, the normal velocity of a thin combustion front may be typically treated as a constant value independent of the large-scale gas flow of the fuel mixture \cite{Zeldovich-et-al-1985,Law-book,Bychkov-Liberman-2000}. On the contrary, the normal velocity of a doping front is determined by the electric field in the undoped material just ahead of the front, see Eq. (\ref{eq1-instab}).
Finally, the last term in Eq. (\ref{eq4-instab-preliminary}) has not been encountered before in  combustion science or any other study of front dynamics. This term is intrinsic for the doping fronts in OSCs, and, as we will see below, it provides additional nonlinear destabilization of the front since it has the same origin as the linear destabilizing term $\hat {J}\tilde {X}_{p}$.  The nonlinear destabilization may be quite strong in the 3D geometry so that the whole approach of weak nonlinearity fails and one has to go back to the fully nonlinear equation Eq. (\ref{M4}), which is much more difficult for solution than  Eq. (\ref{eq4-instab-preliminary}). Still, in the 2D geometry of the OSCs films studied experimentally in Ref. \cite{Bychkov-et-al-2011} the approach of weak nonlinearity provides reasonable results able to explain the experimental data both qualitatively and quantitatively.

In the studies of the unstable p-doping fronts we also have to take into account linear stabilization of the instability at short wavelengths below the cut-off $\lambda_{p}\propto L_{p}$. The short wavelength stabilization is related to the small front thickness and, therefore, was not accounted in Eq.~(\ref{dispersion-relation}). Though negligible at the experimentally observed length scales, the stabilization prevents the ultraviolet divergence of the perturbation spectrum in the doping front dynamics by use of the modified dispersion relation
\begin{equation}
\label{growth rate with cut-off}
\sigma = U_{p}\left( \left| {k} \right| -\frac{ \lambda_{p}k^{2}}{2\pi}\right).
\end{equation}
The above dispersion relation may be also presented in the form of a linear integral-differential equation similar to Eq. (\ref{dispersion-relation-nonlin})
\begin{equation}
\label{linear-instab-with-the-cut-off}
U_{p}^{ - 1}\frac{\partial \tilde {X}_{p}}{\partial t} = \hat {J}\tilde {X}_{p}  + {\frac{{\lambda _{p}}} {{2\pi
}}}\frac{\partial^{2} \tilde {X}_{p}}{\partial y^{2}}.
\end{equation}
The linear stabilization terms have the physical meaning of the local front velocity depending on the front curvature \cite{Zeldovich-et-al-1985,Law-book,Bychkov-Liberman-2000}; these terms may be also incorporated as curvature corrections to the doping front velocity in Eq. (\ref{eq1-instab}).
Then, taking into account the linear stabilization term in the nonlinear equation Eq. (\ref{eq4-instab-preliminary}), we come to the final equation for the p-doping front
\begin{equation}
\label{eq4-instab}
\frac{1}{U_{p}}\frac{\partial \tilde {X}_{p}}{\partial t} = \hat {J}\tilde {X}_{p} + \left[\frac{\partial \tilde {X}_{p}}{\partial y}\right]^{2} + \frac{1}{U_{p}} \hat {H}{\left[ \frac{\partial \tilde {X}_{p}}{\partial t}\frac{\partial \tilde {X}_{p}}{\partial y} \right]} + {\frac{{\lambda _{p}}} {{2\pi
}}}\frac{\partial^{2} \tilde {X}_{p}}{\partial y^{2}}.
\end{equation}
Equation (\ref{eq4-instab}) has some common features with the Sivashinsky equation and other nonlinear equations derived within the theory of the DL
hydrodynamic flame instability \cite{Bychkov-Liberman-2000,Sivashinsky,Zhdanov-Trubnikov-1989,Joulin-1991,Bychkov-1998}.
Due to this similarity, we may employ the mathematical methods of investigating the DL instability to the  problem of doping fronts in OSCs as demonstrated in the next section.

\section{Shape and velocity of stationary curved p-doping fronts}

Here we consider stationary periodic solutions
to the derived nonlinear equation Eq. (\ref{eq4-instab}) in the form $\tilde{X}_{p}(y,t)=t\Delta U + f (y)$,
where $\Delta U$ is additional increase of the doping front velocity because of the instability and $f(y)$ is a periodic function of the period $\lambda$. Despite the imposed limitations, the stationary periodic solutions provide the clue to the general understanding of the p-doping front dynamics.
 We reduce Eq. (\ref{eq4-instab}) to the stationary form
\begin{equation}
\label{eq4-instab-stat}
\Delta U/U_{p}= \left(1+\Delta U/U_{p}\right) \hat {J} f + \left(\frac{\partial f}{\partial y}\right)^{2} + {\frac{{\lambda _{p}}} {{2\pi
}}}\frac{\partial^{2} f}{\partial y^{2}}.
\end{equation}
We point out that $\hat {J} f$ is the key term describing the front instability at the linear stage; thus the product $(\Delta U/U_{p}) \hat {J} f$ with $\Delta U>0$ means additional nonlinear destabilization of the front. It should be stressed that such a term has not been encountered before in the combustion studies of the DL instability, and it modifies the nonlinear dynamics of the p-doping fronts quite strongly as compared to the combustion fronts.

By employing the method of pole decomposition \cite{Joulin-1991,Bychkov-1998,Vaynblat-Matalon-2000,Karlin-Makhviladze-2003,Denet-2007}, we look for $\lambda $-periodic solutions to Eq. (\ref{eq4-instab-stat}) in the form
\begin{equation}
\label{f-pole}
f\left( y \right)=\frac{{{\lambda }_{p}}}{2\pi }\sum\limits_{i=1}^{2N}{\ln \sin \left[ \frac{k}{2}\left( y-{{y}_{\alpha }} \right) \right]},
\end{equation}
where the complex values $y_{\alpha}$  represent the poles, unknown so far. Substituting the pole solution Eq. (\ref{f-pole}) to the stationary equation (\ref{eq4-instab-stat}) we derive the set of equations for the poles $y_{\alpha}$. In these calculations,  action of the DL operator $\hat {J}$ on Eq. (\ref{f-pole}) requires additional comments. The definition of the operator $\hat {J}$ may be re-written as follows
\begin{equation}
\label{DL-operator2}
\hat{J}\left[f\left(y\right)\right]=\frac{2}{\lambda}\sum\limits_{n=1}^{\infty}
{kn\int\limits_{0}^{\lambda}{\cos\left[kn\left(y-y'\right)\right]}f\left(y'\right)dy'},
\end{equation}
where the integral in Eq. (\ref{DL-operator2}) is similar to the Fourier transform.
We  integrate Eq. (\ref{DL-operator2}) by parts  using the following identities \cite{Gradshteyn-Ryzhik}
\begin{equation}
\label{identity_1}
\frac{\sin{y}}{\cosh{z}-\cos{y}}=2\sum\limits_{n=1}^{\infty}{e^{-nz}\sin{ny}},\;\;z>0,
\end{equation}
\begin{equation}
\label{identity_2}
2\sum\limits_{n=1}^{\infty}{e^{-nz}\cos{ny}}=\frac{\sinh{z}}{\cosh{z}-\cos{y}}-1,\;\;z>0.
\end{equation}
As a result, we can specify action of the DL operator on the pole solution Eq. (\ref{f-pole}) as
\begin{equation}
\label{DL-operator3}
\hat{J}f=k-\frac{k\sinh{\left[k{\varepsilon}_{\alpha }\textrm{Im}{\left({y}_{\alpha }\right)}\right]}}{\cosh{\left[k{\varepsilon}_{\alpha }\textrm{Im}{\left({y}_{\alpha }\right)}\right]}-\cos{\left[k\left(y+\textrm{Im}{\left(y_{\alpha}\right)}\right)\right]}},
\end{equation}
with the  designation  ${\varepsilon}_{\alpha }=\textrm{sign}\left[ \textrm{Im}{\left(y_{\alpha}\right)}\right]$ introduced for brevity.
Then, we substitute Eq. (\ref{f-pole}) to Eq. (\ref{eq4-instab-stat}) and perform  straightforward calculations to find the set of equations for the poles
\begin{equation}
\label{pole-equations-1}
\frac{\lambda_{p}/\lambda}{ 1+{\Delta U}/{{{U}_{p}}}} \sum_{\alpha \neq \beta} ^{2N-1} \cot \left[\pi \frac{y_{\alpha}-y_{\beta}}{\lambda}\right] + i \textrm{sign}(\textrm{Im} y_{\alpha})=0,
\end{equation}
where $N$ pairs of poles are involved in the solution.
In the derivation of Eq. (\ref{pole-equations-1}) we also employed the following trigonometric formula
\begin{equation}
\cot a \cot b=\cot (a-b)\left[\cot b - \cot a\right]-1.
\end{equation}
Apart from the equations for the poles, the calculations yield  an equation for the velocity increase
\begin{equation}
\label{velocity-increase}
\frac{\Delta U / {U}_{p}}{{1+\Delta U/{U}_{p}}}=N\frac{{{\lambda }_{p}}}{\lambda }\left( 1-\frac{{N {\lambda }_{p}/\lambda}}{ 1+\Delta U/{U}_{p}} \right),
\end{equation}
where $N$  is an integer standing for the number of the pole pairs.
Solution to Eq. (\ref{velocity-increase}) is illustrated in Fig. 4 (a) for $N=1$: the solid lines present the solutions with  velocity increase limited for any ${\lambda }_{p}/\lambda$; the dashed lines show the solutions with unlimited velocity increase. The latter, presumably, does not have a physical meaning and hence will be omitted in the following.
The solution presented by the solid line is determined by a simple formula
\begin{equation}
\label{velocity-increase-2}
\Delta U /U_{p}= N \lambda_{p}/\lambda \quad \textrm{for} \quad  N \le \lambda / \lambda_{p},
\end{equation}
which follows from Eq. (\ref{velocity-increase}). We can also obtain the same result for the velocity increase by taking the spatial averaging of   Eq. (\ref{eq4-instab-stat}) as
\begin{equation}
\label{velocity-average}
\frac{\Delta U}{{{U}_{p}}}=\frac{1}{\lambda }\int\limits_{0}^{\lambda}{{{\left( \frac{\partial f}{\partial y} \right)}^{2}}dy}.
\end{equation}
\begin{figure}
\centering
\includegraphics[width=3.3in]{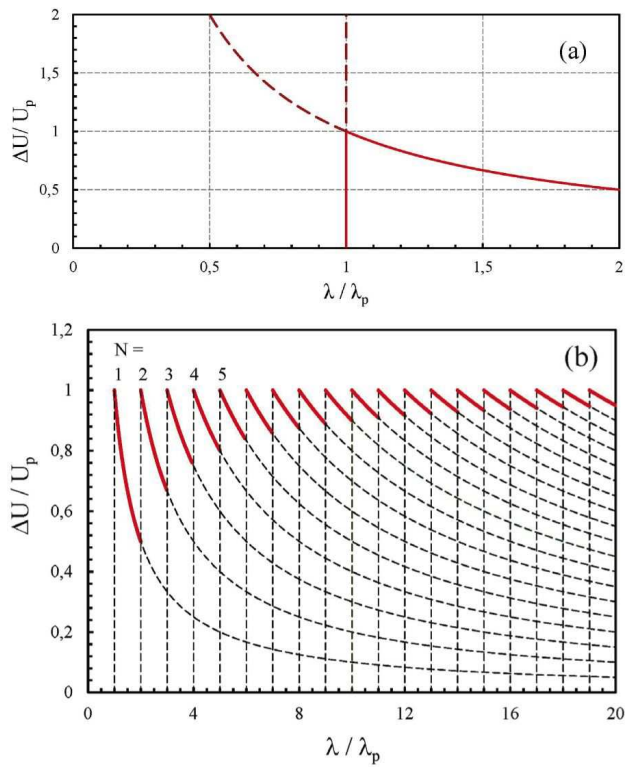}
\caption{Velocity increase $\Delta U /U_{p}$ of the p-doping front for stationary periodic solutions versus the dimensionless period $\lambda / \lambda_{p}$: (a) for one pair of poles $N=1$; the solid line indicates the limited solution, the dashed line indicates the diverging solution; (b) for different numbers of pole pairs $N$; the solid lines indicate maximal possible velocity increase for a fixed period, the dashed lines present other  solutions.}
\end{figure}
The velocity increase is presented in Fig. 4 (b) versus the scaled period of the solution $\lambda/\lambda_{p}$ for different numbers of the pole pairs. As we can see from Fig. 4 (b), a curved p-doping front is possible only for a sufficiently large period $\lambda$ exceeding the instability cut-off $\lambda \geq \lambda_{p}$; the branch starting at $\lambda = \lambda_{p}$ corresponds to one pair of poles, $N=1$. The solution with $N$ pairs of poles becomes possible for $\lambda > N\lambda_{p}$ and it provides the maximal front velocity  within the interval of $N < \lambda / \lambda_{p} < N+1$.
As we increase the period $\lambda$ in comparison with $\lambda_{p}$, multiple stationary periodic solutions of the same period become possible with different numbers of the pole pairs. In Fig. 4 (b), the solid lines indicate the solutions corresponding to the maximal velocity increase for a fixed period $\lambda$; other solutions are presented by the dashed lines. The solutions with maximal velocity involve also the largest possible number of poles for a fixed period. Presumably, the solutions with the maximal propagation velocity and the maximal number of poles are stable, while other solutions (shown by the dashed lines) are unstable. In the context of combustion fronts, such a result has been demonstrated in Ref. \cite{Vaynblat-Matalon-2000} by investigating stability of the pole solutions; we expect the same conclusion to hold for the doping fronts in OSCs as well, though the detailed stability analysis requires much work and is left for the future. Figure 4 (b) demonstrates also that the maximal velocity increase tends to the limiting value
\begin{equation}
\label{velocity-max}
\textrm{max} \Delta U /U_{p} = 1
\end{equation}
as the period $\lambda$ becomes much larger than the cut-off $\lambda_{p}$, i.e. for $\lambda/\lambda_{p}\rightarrow\infty$. The limit of large scales in comparison with the cut-off is of the most importance for the analysis of the experimental results   \cite{Bychkov-et-al-2011}. Thus, we expect strongly curved p-doping fronts in OSC films to propagate twice as fast in comparison with the planar fronts.

Next, by using Eq. (\ref{pole-equations-1}), we investigate the front shape predicted by the stationary solutions. Translational invariance of the solution Eq. (\ref{f-pole}) implies that we can chose arbitrary $\textrm{Re}(y_{\alpha})$; hence we set $\textrm{Re}(y_{\alpha})=0,$ so that the poles $y_{\alpha}$ become pure imaginary  $y_{\alpha}=\pm iB_{\alpha}$. Taking into account complex conjugate pairs, we rewrite Eq. (\ref{f-pole}) in a more  convenient form
\begin{equation}
\label{f-pole2}
f\left( y \right)=\frac{{{\lambda }_{p}}}{2\pi }\sum\limits_{i=1}^{N}{\ln \frac{1}{2}\left[ \cosh \left( k{{B}_{i}} \right)-\cos \left( ky \right) \right]}
\end{equation}
and the set of equations for the poles, Eq. (\ref{pole-equations-1}), reduces to
\begin{align}
\label{pole-equations-2}
  &\frac{\lambda }{\lambda_{p}}\left( 1+\frac{\Delta U}{{U}_{p}} \right)=\coth \left[ k  B_n  \right] +  \\ \nonumber
  &\sum\limits_{n\ne m}^{N}{\left\{ \coth \left[ \frac{k}{2}\left( {{B}_{n}}+ {{B}_{m}} \right) \right]+ \coth \left[ \frac{k}{2}\left( {{B}_{n}}- {{B}_{m}} \right) \right] \right\}},
\end{align}
for $n=1,\ldots N$.
Figure 5 (a) shows positions for the poles $B_{\alpha}$ calculated for  Eq. (\ref{pole-equations-2}) for different number of pole pairs $N$ and the solution period  $\lambda / \lambda_{p}=N$ corresponding to the maximal velocity increase. As we can see in Fig. 5 (a), scaled positions of the poles $B_{\alpha}$ decrease strongly for any fixed $\alpha$ as the number of poles increases. Since the solution (\ref{f-pole2}) involves linear superposition of the poles, then we can study relative roles of different pole pairs separately. As an illustration, Fig. 5 (b) presents the solution (\ref{f-pole2}) with three pairs of poles for $\lambda / \lambda_{p}=N=3$ together with separate contributions from different poles $\alpha = 1 - 3$ (marked in Fig. 5 (a)). We observe in the figure that the pole pair with the minimal number $\alpha = 1$ plays the dominant role in building the amplitude of the final  hump, while the poles with higher numbers are  responsible mostly for fine details. In that sense, the role of poles in the solution (\ref{f-pole2}) may be compared qualitatively to the role of harmonics in Fourier series.

\begin{figure}
\centering
\includegraphics[width=3.3in]{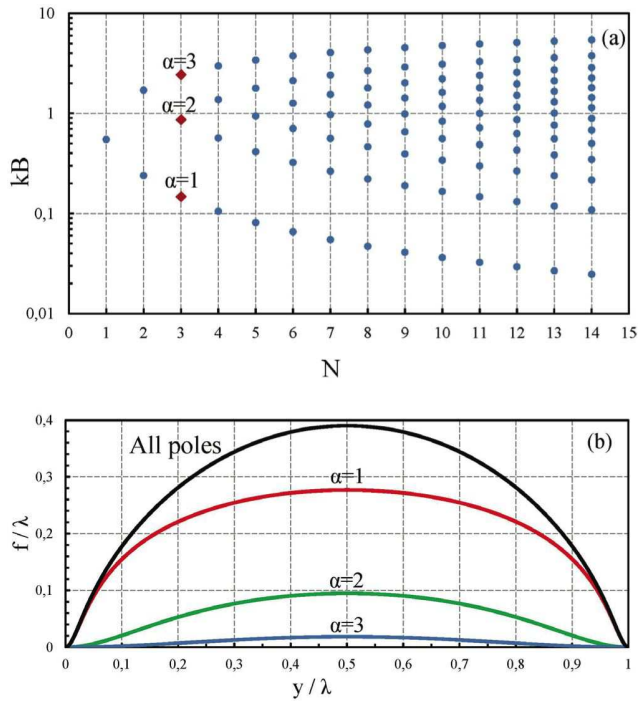}
\caption{Illustration of the pole solutions: (a) Scaled pole positions $B_{\alpha}$ versus the number of pole pairs for $N=\lambda /\lambda_{p}$; (b) Solution for three pole pairs for $\lambda /\lambda_{p}=N=3$ and separate contributions of different pairs, $\alpha = 1,2,3$ selected in the plot (a).}
\end{figure}

Substituting the solutions for the poles into Eq. (\ref{f-pole2}) we find the shapes of the  curved p-doping fronts. In agreement with the experimental observations, Fig. 1, the shapes are built of smooth humps facing the undoped material and sharp cusps pointing to the doped one, e.g. see Fig. 6 (a) with the solution plotted for $\lambda/\lambda _{p}=2$. For illustrative purposes, Fig. 6 (a) presents three periods of the solution. We remind that in this section we focus on the stationary  solutions with humps and cusps reproduced periodically. Still, these humps provide the key to understanding  irregular front patterns as well like those observed experimentally and shown in Fig. 1.  Figure 6 (b) presents the stationary humps corresponding to the maximal front velocity and the maximal number of poles found for different values of the scaled period $\lambda/\lambda _{p} =N=1-10$. In Fig. 6 (b) we observe only minor variations of the hump shape with increasing period. The variations concern mostly the cusps, which become sharper as we increase the hump size $\lambda$ in comparison with the instability cut-off $\lambda _{p}$. We remind that the limit of large length scales, $\lambda/\lambda_{p}\rightarrow\infty$, is the most interesting from the experimental point of view.
As we can see in Fig. 6 (b),  the shape of a hump becomes self-similar in the limit of large length scales, and this shape may be approximated with a very good accuracy by a solution with  a limited number of pole pairs.  In particular,  Fig. 6 (b) demonstrates negligible difference between the solutions obtained for  $\lambda/\lambda _{p} =N=5$ and $10$.
\begin{figure}
\centering
\includegraphics[width=3.5in]{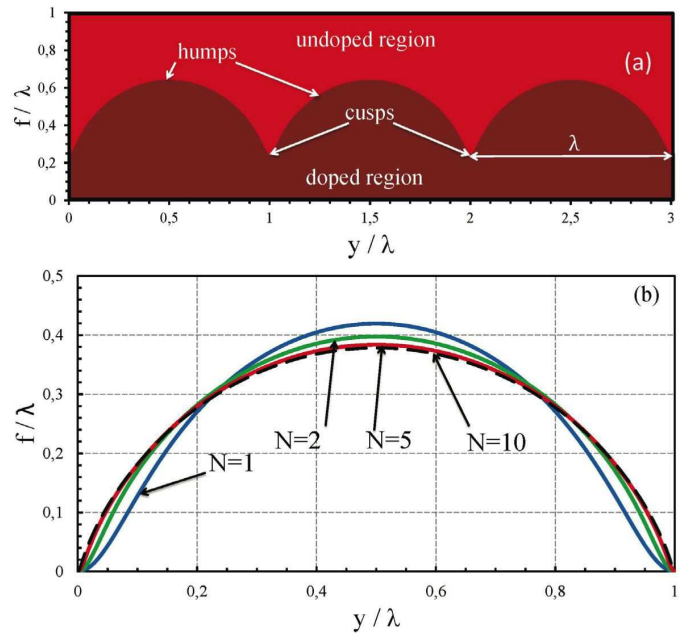}
\caption{Shapes of the curved p-doping fronts: (a) periodic solution plotted for $\lambda/\lambda _{p}=2$.
(b) Shapes of humps for $\lambda/\lambda _{p}=N=1,2,5,10$.}
\end{figure}

At the end of this section we point out that the pole solutions may be written in an explicit analytical form for small numbers of pole pairs $N=1,2$.
For one pair of poles   ($N=1$)  we find from Eq. (\ref{pole-equations-2})
\begin{equation}
\label{one_pole}
\frac{\lambda }{\lambda_{p}}\left( 1+\frac{\Delta U}{{U}_{p}} \right)=\coth \left[ k  B_n  \right],
\end{equation}
which leads to the front shape
\begin{equation}
f=\frac{{{\lambda }_{p}}}{2\pi }\ln \left[ 1-\sqrt{1-\left( \frac{{\lambda}_{p}}{\lambda+{\lambda}_{p}} \right)^{2}}\cos \left( \frac{2\pi y}{\lambda } \right) \right].
\end{equation}
In the same way, for two pairs of poles ($N=2$) we obtain
\begin{eqnarray}
\label{two_pole}
&&k{{B}_{1,2}}=\text{arctanh}\left[ \frac{3\lambda_p/\lambda}{1+\Delta U/U_p} \right]\mp \\ \nonumber
&&\frac{1}{2}\text{arccosh}\left\{\left[ 1-{{\left( \frac{3\lambda_p/\lambda}{1+\Delta U/U_p} \right)}^{2}} \right]^{-1}\right\}
\end{eqnarray}
and the front shape
\begin{eqnarray}
&&f=\frac{\lambda _p}{2\pi }\ln \frac{1}{4}\left[ \frac{1+\Lambda/2}{ 1-\Lambda }- \right.\\
\nonumber
&&\left.\sqrt{2}\frac{\sqrt{2-\Lambda }}{1-\Lambda}\cos \left( \frac{2\pi y}{\lambda } \right) +{{\cos }^{2}}\left( \frac{2\pi y}{\lambda } \right) \right],
\end{eqnarray}
where the designation $\Lambda = \left[3\lambda_p/(\lambda+2\lambda_p) \right]^{2}$ has been introduced for brevity.

\section{Comparison to the experiments and technical implications}

In this Section we compare the theoretical predictions to the experimental results of Ref. \cite{Bychkov-et-al-2011} and demonstrate how theoretical understanding  may be employed for better control and optimization of the electrochemical doping process in optoelectronic devices. In particular, Figure 7 compares positions of the doping fronts as predicted by the theory and measured experimentally. Here, the solid lines present front positions assuming the planar front shape (labeled "theory 1") and accounting for the instability and the curved shape of the p-doping front (labeled "theory 2").
The markers show average positions of the strongly corrugated fronts obtained experimentally in Ref. \cite{Bychkov-et-al-2011} and illustrated in Fig. 1; the error-bars indicate average spreading of the front brush.
Details of the theoretical calculations for the geometry of the planar fronts may be found in Ref. \cite{Modestov-et-al-2011}. Here we remind  two important features of the process taken into account in the calculations: (i) according to the experiments \cite{Bychkov-et-al-2011}, the n-doping front starts later and propagates noticeably slower than the p-doping front; (ii) the fronts accelerate towards each other because of much higher conductivity of the doped material in comparison with the undoped one. Due to the relatively high conductivity of the doped material, the potential difference applied to the electrodes drops effectively at the  gap between the fronts. As the distance between the p- and n-fronts decreases,  the electric field in the gap increases, which leads to increase of the front velocities according to Eq. (\ref{eq1-instab}). Theoretically, the electric field and the planar front velocities diverge as the fronts meet; the front propagation comes to a complete halt at that point.
The theory of Sec. IV predicts that, in the experimental conditions, the doping front instability and the corrugated shape may increase the propagation velocity of the p-doping front by a factor of 2. We employed this theoretical result for calculating the average  position of the corrugated p-doping front labeled as "theory 2" in Fig. 7. Since the theory is valid only for p-doping fronts, then we use the model of a planar n-doping front as in the previous calculations. Still, the n-doping front sweeps only about 10\% of the active material in the described experiments, and, therefore, a particular model for the n-front dynamics produces only slight modifications of the results.
\begin{figure}
\centering
\includegraphics[width=3.3in]{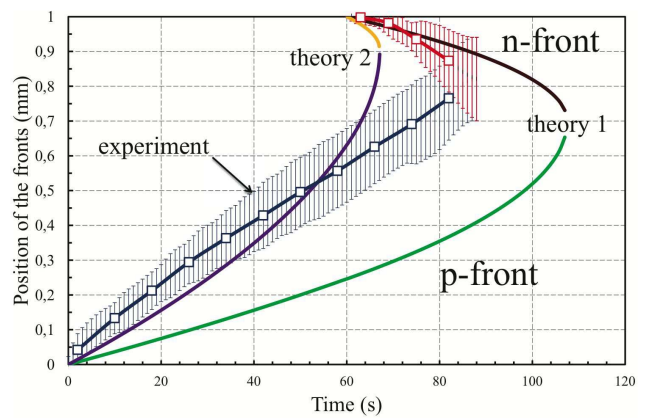}
\caption{Position of the doping fronts versus time as predicted by the theory for the planar fronts (theory 1) and taking into account the instability (theory 2). The experimental results of Ref. \cite{Bychkov-et-al-2011} for the average front positions are presented by the markers. The error bars indicate the average spreading of the front brush.}
\end{figure}

Figure 7 shows that the well-developed front instability and the curved shape of the p-doping front may speed-up the doping process considerably thus improving   performance of the optoelectronic devices. The theoretical predictions for the p-front dynamics are in a good agreement with the experimental data during most part of the doping process (for the time instants up to $\sim 65 \, \textrm{s}$).
 The front velocity found experimentally  is only a little lower than the theoretical expectations for the maximal front velocity within that interval. Some deviations in the front position and velocity between the theory and the experiments observed in Fig. 7 are quite natural, since the periodic solutions studied theoretically in Sec. IV are not identical to the irregular pattern of Fig. 1 obtained experimentally.
 As we discuss below, the difference in the front shape should lead to  larger velocity of the periodic curved fronts of Sec. IV in comparison with the  front pattern of Fig. 1. The deviations between the theory and the experiments become considerable at the end of the process, for the time instants about $(70-80) \, \textrm{s}$. On the other hand,  the very approach of the average front position is misleading at the end of the doping process, since the gap between the p- and n-fronts becomes  comparable to the size of the humps.
 \begin{figure}
\centering
\includegraphics[width=3.3in]{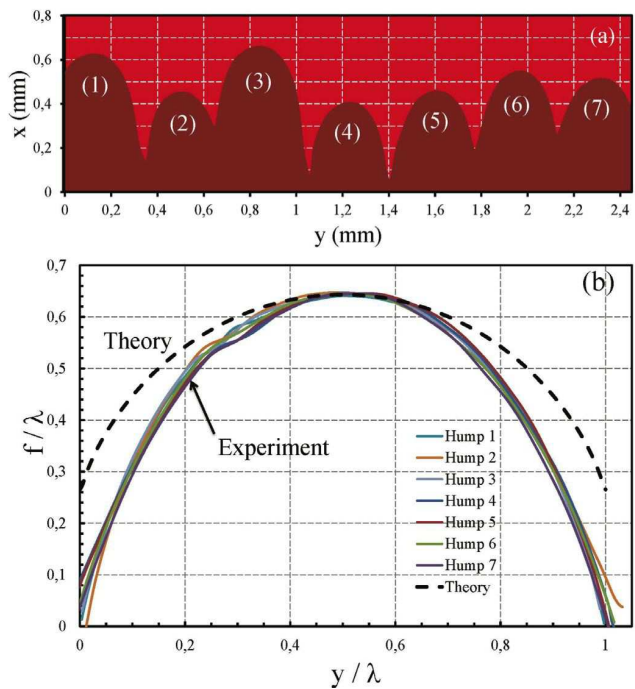}
\caption{Shape of  the  humps obtained in the experiments \cite{Bychkov-et-al-2011} for the time instant of $54\, \textrm{s}$, (a), and comparison of the experimental results for different humps to the theoretical predictions  for a stationary periodic solution, (b). The experimental results in the plot (b) are shifted and re-scaled in order to demonstrate self-similarity of the humps.}
\end{figure}

 Figure 8 compares  shape of the humps at the curved p-doping front as predicted by the theory for a stationary periodic solution and obtained in the experiments \cite{Bychkov-et-al-2011}. In order to perform the comparison, we shift and re-scale the humps  labeled in Fig. 8 (a) keeping equal scales in both directions. We point out that comparison is possible only for the top part of the humps, since the bottom parts are strongly affected by particular irregular conditions of front initiation. Figure 8 demonstrates a reasonable agreement between the theory and the experiments. Still, we can see that the periodic stationary humps studied in Sec. IV are somewhat smoother and the cusps in between them are more shallow than  those observed in the experiments. In part, the difference may be attributed to the limited accuracy of the assumption of weak nonlinearity used for the derivation of Eq. (\ref{eq4-instab}) from the fully nonlinear equation Eq. (\ref{M4}). However, the fully nonlinear equation Eq. (\ref{M4}) is much more difficult for the analysis than Eq. (\ref{eq4-instab}); for this reason, investigation of Eq. (\ref{M4})  is left for future studies. Large amplitude of the humps in Fig. 1 may also originate in a particular way of initiating the instability in the experiments \cite{Bychkov-et-al-2011}. As demonstrated in Fig. 1, the strong instability of the p-doping front has been initiated by introducing minor corrugations at the anode surface. Because of the local increase of the electric field at the corrugations, the doping process started faster at the corrugation tips. For a while, the regions of p-doping developed separately in the form of isolated humps close to the anode corrugations without forming a single continuous p-front. When the front was finally formed, the employed way of initiating the instability produced extremely deep cusps -- chasms -- in between the humps, which remained quite deep till the end of the doping process thus reducing the average front position and the average front speed. Here we would like to stress an important difference between the doping fronts in OSCs and combustion fronts one more time; the  similarity between these two phenomena is limited and may be often misleading.
 In the case of combustion, normal velocity of the front propagation is approximately constant, so that deep chasms at a combustion front imply strong increase of the front speed, and the Huygens stabilization mechanism leads to fast leveling of the chasms as shown in Fig. 3. In contrast to combustion, such effects do not happen at the doping fronts in OSCs. The doping front velocity is determined by the electric field just ahead of the front, which is strongly reduced in the chasms.
 In particular, Fig. 9 presents the electric field (a) and potential (b) calculated for the corrugated doping front of Fig. 8 (a). The electric field in Fig. 9 varies by $3-4$ orders of magnitude from $\sim 12 \textrm{kV/m}$ at the tips of the humps to $\sim (10 - 100) \textrm{V/m}$ at the bottom of the chasms.
\begin{figure}
\centering
\includegraphics[width=3.2in]{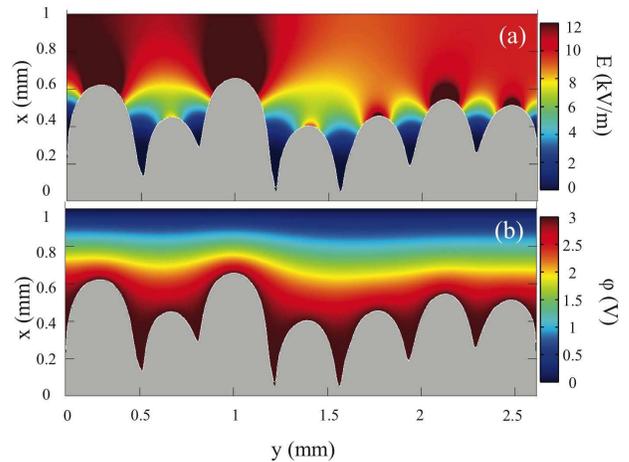}
\caption{Absolute value of the electric field (a) and potential (b) calculated in the undoped region for the corrugated p-doping front of Fig. 8 (a).}
\end{figure}
\begin{figure}
\centering
\includegraphics[width=3.2in]{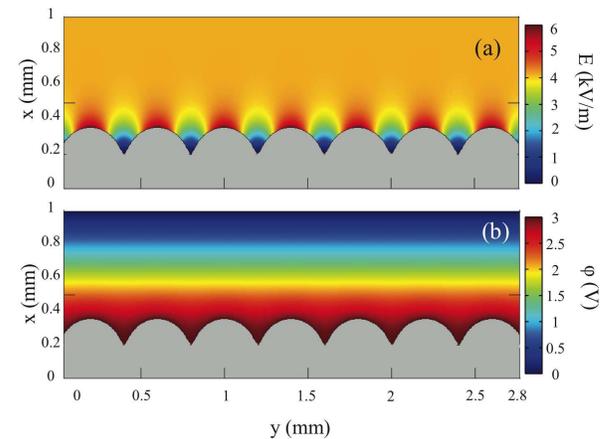}
\caption{Absolute value of the electric field (a) and potential (b) calculated in the undoped region for the moderately corrugated p-doping front obtained in Sec. IV with the same number of humps as in Fig. 9.}
\end{figure}
For this reason, the Huygens stabilization mechanism is much weaker for the doping fronts, and deep chasms reduce the average front velocity instead of increasing it. As a result, moderately corrugated doping fronts studied in Sec. IV provide higher average propagation velocity than the strongly corrugated fronts of Fig. 1. The electric field (a) and potential (b) in the undoped material ahead of a moderately corrugated doping front studied in Sec. IV is shown in Fig. 10; in that case the electric field between the humps and the cusps varies much less than in Fig. 9, approximately by an order of magnitude, from $\sim 5 \textrm{kV/m}$ to $\sim 0.5 \textrm{kV/m}$. Thus, in order to obtain further increase of the doping rate in comparison with the experiments \cite{Bychkov-et-al-2011}, one should avoid extremely deep chasms when initiating the doping front instability. This can be achieved, for example, by introducing smoother corrugations at the electrodes imitating the curved periodic doping fronts obtained in Sec. IV.

 Another important feature of the process is the self-similar shape of the humps at the corrugated p-doping front, which allows additional control of the front corrugations and, hence, of the final light-emitting p-n junction curve. As we have shown in Sec. IV, in the limit of large length scales $\lambda / \lambda_{p} \gg 1$ inherent to the experiments \cite{Bychkov-et-al-2011}, the maximal velocity increase tends to the limiting value $\Delta U / U_{p}\rightarrow 1$ independent of a particular length scale $\lambda$, and the  humps attain a self-similar shape presented in Fig. 6. We point out that the sample sizes  in the experiments \cite{Bychkov-et-al-2011},  namely, breadth of $1$ mm and length of $1$ cm, were many orders of magnitude larger than characteristic width of the doping front, $L_{p} \sim (10^{ - 4}{\rm -}  10^{ - 3}) {\rm mm}$, determined by ionic diffusion \cite{Modestov-et-al-2010,Modestov-et-al-2011}.
 For this reason, humps of both large and small size in comparison with the sample sizes may satisfy the limit $\lambda \gg \lambda_{p}  \propto L_{p}$ and provide maximal velocity of front propagation. As an illustration, Fig. 11 shows two p-doping fronts with two different periods of the humps, $\lambda=0.3 \,{\rm mm}$   and $1.2 \,{\rm mm}$, which propagate with the same maximal velocity. Visually, front (b) demonstrates  strong  corrugations, while front (a) is almost planar in average when compared to the sample sizes. Respectively, in the case (b) one should expect a strongly curved light-emitting p-n junction, while in the case (a) the p-n junction will be almost planar with slight wrinkles, though both cases demonstrate the same maximal doping front velocity increased because of the instability.
   \begin{figure}
\centering
\includegraphics[width=3.3in]{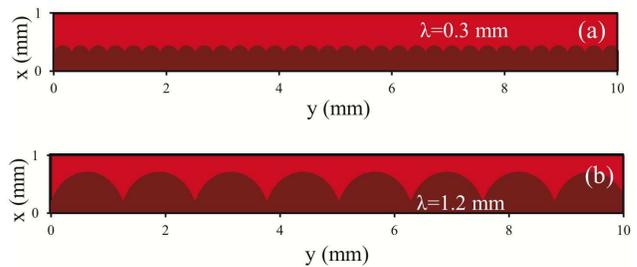}
\caption{P-doping fronts with small-scale (a) and large-scale (b) corrugations as compared to the sample size, with periods $\lambda = 0.3 \textrm{mm}$ and $1.2 \textrm{mm}$, respectively. Both fronts propagate with the same maximal velocity.}
\end{figure}

Finally, we discuss implications of the present analysis for the 3D geometry of the doping fronts. Though the experiments of Ref. \cite{Bychkov-et-al-2011} concerned only 2D OSC films, the 3D bulk geometry of the doping process is also technically attainable for transparent electrodes made, e.g., of chemically derived graphene \cite{Matyba-et-al-2010}. In the 3D geometry, the stationary weakly nonlinear equation (\ref{eq4-instab-stat}) becomes
\begin{equation}
\label{eq4-instab-stat-3D}
\Delta U/U_{p}= \left(1+\Delta U/U_{p}\right) \hat {J} f + \left(\nabla f \right)^{2} + {\frac{{\lambda _{p}}} {{2\pi
}}}\nabla^{2} f,
\end{equation}
where  $\nabla$-operator concerns two transverse coordinates, $y$ and $z$. By using the methods of combustion theory \cite{Bychkov-Liberman-2000,Denet-2007,Blinnikov-Sasorov-1996}, we may look for a particular solution to Eq. (\ref{eq4-instab-stat-3D}) in the form
\begin{equation}
\label{3D-solution}
f_{3D}(y,z) = f_{2D}(y) + f_{2D}(z).
\end{equation}
The solution (\ref{3D-solution})  describes only rectangular 3D patterns at the doping fronts as illustrated in Fig. 12. Still, comparison of triangular, rectangular,  hexagonal and axisymmetric patterns for a similar combustion problem demonstrated  that increase of the front velocity depends slightly on a particular pattern   \cite{Bychkov-Liberman-2002}. The combustion studies suggest also the evaluation $\textrm{max}(\Delta U_{3D})\approx 2\, \textrm{max}( \Delta U_{2D})$, which would imply velocity increase by a factor of 3 for curved  p-doping fronts in the 3D bulk OSC devices. However, careful investigation of Eq. (\ref{eq4-instab-stat-3D}) indicates that velocity increase of a curved p-doping front with 3D patterns is even stronger than that suggested by the above estimate. Indeed, by substituting the  solution Eq. (\ref{3D-solution}) into the stationary equation (\ref{eq4-instab-stat-3D}), we come to the following equation for the maximal velocity increase in the 3D geometry
\begin{equation}
\label{velocity-increase-3D}
\textrm{max}\left[\frac{\Delta U_{3D} / {U}_{p}}{(1+\Delta U_{3D}/{U}_{p})^{2}}\right]=\frac{{1}}{2 },
\end{equation}
which looks similar to Eq. (\ref{velocity-increase}), but, surprisingly, has no physical solution. The last fact implies some important consequences for curved doping fronts in optoelectronic devices with the 3D bulk geometry: (i) the approach of weak nonlinearity of Eq. (\ref{eq4-instab}) does not work in the 3D geometry, and one has to employ the fully nonlinear equation (\ref{M4}); (ii) presumably, the  doping front instability is so strong in the bulk geometry, that it cannot be not stabilized  at the nonlinear stage to any stationary/quasi-stationary solution. Instead of the stabilization, one may expect acceleration of the 3D humps at the doping fronts resembling  the breakdown in lightning   and development of streamers \cite{Raizer-et-al-2010,Milikh-Shneider-2008}. Such an acceleration of the doping fronts with 3D patterns may be also compared to  acceleration of curved  combustion fronts in deflagration-to-detonation transition
\cite{Bychkov-et-al-2005,Bychkov-et-al-2008}. Similar to the present work, combustion studies have encountered an example of curved flames in channels with the propagation regime determined by the geometry dimension: in open tubes, the combustion fronts propagate stationary in average in the 2D geometry, but accelerate in the 3D case, see \cite{Akkerman-et-al-2006,Akkerman-et-al-2010}. Thus, one should expect new interesting effects for doping front dynamics in the 3D bulk optoelectronic devices, which are qualitatively different from the case of 2D OSC films, and which may further improve the device kinetics.
  \begin{figure}
\centering
\includegraphics[width=3.4in]{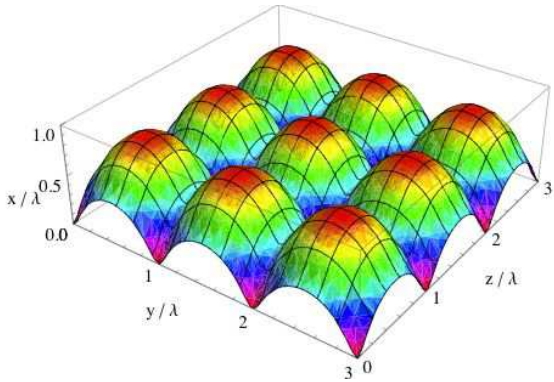}
\caption{A corrugated p-doping front with a rectangular 3D pattern.}
\end{figure}

\section{Summary}

In the present work we have developed a nonlinear theory of the doping front instability discovered recently in Ref.\ \cite{Bychkov-et-al-2011}. Our nonlinear studies predict the possibility of quasi-stationary curved p-doping fronts in OSC films, such as the ones employed, e.g., in light-emitting cells. The curved doping fronts may propagate with a velocity that is increased by a factor of 2 in comparison with planar fronts, thus allowing for an optimization of the optoelectronic devices. The theoretical predictions for the velocity increase are in good agreement with the experimental measurements, see e.g.\ Ref. \cite{Bychkov-et-al-2011}. According to the theory, further increase of the front velocity may be achieved by fine tuning the initiation of the instability using minor smooth humps introduced at the electrodes.

The theory also predicts the front shape in the form of smooth humps facing the undoped material and sharp cusps facing the doped substance, in agreement with experimental observations. At the same time, in contrast to analogous studies of combustion front dynamics, extra deep cusps at the doping fronts do not produce additional increase of the front velocity; instead, they lead to certain velocity reduction in comparison with the moderately corrugated doping fronts. In the experimentally important parameter domain, humps at the p-doping fronts demonstrate self-similar properties, which allows for better control of the front shape and the doping rate. In particular, we suggest that a maximal doping rate may be achieved both for strongly corrugated fronts, and for  fronts which look  planar on average on large scales but demonstrate corrugations on small scales. This interesting property of the doping front instability allows control of the shape of the light-emitting p-n junction.

Finally, the present analysis suggests that dynamics of the curved p-doping fronts in optoelectronic devices with a 3D bulk geometry may be qualitatively different from that in the 2D OSC films studied in this paper. We expect that p-doping fronts with 3D patterns will not be stabilized to a stationary (or quasi-stationary) nonlinear pattern even in the case of a single front propagating in an initially uniform external electric field, but they will rather accelerate,  similar to breakdown in lightning. Such an acceleration may reduce the turn-on time in the 3D optoelectronic devices considerably in comparison with the light-emitting cells based on 2D OSC films.

\acknowledgements
The authors thank Piotr Matyba, Ludvig Edman and V'acheslav Akkerman for useful discussions.
The work was supported in part by the Swedish Research Council.

\end{document}